\documentclass[12pt]{article}
\usepackage{epsfig}
\usepackage{graphicx}
\usepackage{a4}
\usepackage{latexsym}
\usepackage{cite}

\usepackage{pslatex}
\usepackage[latin1]{inputenc}
\usepackage[T1]{fontenc}

\newcommand{\hspn}{{\hspace{-4.5mm}}}
\newcommand{\gsim}{\raisebox{-0.07cm}{$\:\:\stackrel{>}{{\scriptstyle
 \sim}}\:\: $} }
\newcommand{\lsim}{\raisebox{-0.07cm}{$\:\:\stackrel{<}{{\scriptstyle
 \sim}}\:\: $} }
\newcommand{\beq}{\begin{equation}}
\newcommand{\eeq}{\end{equation}}
\newcommand{\bea}{\begin{eqnarray}}
\newcommand{\eea}{\end{eqnarray}}
\newcommand{\nn}{\nonumber}
\newcommand{\MSb}{$\overline{\mbox{MS}}$}
\newcommand{\as}{\alpha_{\rm s}}
\newcommand{\ra}{\rightarrow}

\newcommand{\ar}{a_{\rm s}}
 
\def\z#1{{\zeta_{#1}^{}}}
\def\ca{{C^{}_A}}
\def\cf{{C^{}_F}}
\def\nf{{n^{}_{\! f}}}
\def\dabcnc{{{d^{abc}d_{abc}}\over{n_c}}}

\def\cfs{{C_{F}^{\,2}}}
\def\cfa{{C_{FA}}}
\def\cfas{{C_{FA}^{\,2}}}
\def\nfs{{n^{\:\!2}_{\! f}}}
\def\zs#1{{\zeta_{#1}^{\,2}}}
\def\zt#1{{\zeta_{#1}^{\,3}}}

\def\pqq(#1){p_{\rm{qq}}(#1)}
\def\H(#1){{\rm{H}}_{#1}^{}}
\def\Hh(#1,#2){{\rm{H}}_{#1,#2}^{}}
\def\Hhh(#1,#2,#3){{\rm{H}}_{#1,#2,#3}^{}}
\def\Hhhh(#1,#2,#3,#4){{\rm{H}}_{#1,#2,#3,#4}^{}}

\begin{document}
\setlength{\parskip}{0.2cm} \setlength{\baselineskip}{0.55cm}

\begin{titlepage}
\noindent
DESY 07-048  \hfill {\tt arXiv:0708.3731 [hep-ph]}\\
SFB/CPP-07-13 \\
LTH 756 \\[1mm]
August 2007 \\
\vspace{1.5cm}
\begin{center}
\LARGE {\bf Differences between charged-current } \\[2mm]
\LARGE {\bf coefficient functions} \\
\vspace{2.2cm}
\large
S. Moch$^{\, a}$, M. Rogal$^{\, a}$ and A. Vogt$^{\, b}$ \\
\vspace{1.4cm}
\normalsize
{\it $^a$Deutsches Elektronensynchrotron DESY \\
\vspace{0.1cm}
Platanenallee 6, D--15738 Zeuthen, Germany}\\
\vspace{0.5cm}
{\it $^b$Department of Mathematical Sciences, University of Liverpool \\
\vspace{0.1cm}
Liverpool L69 3BX, United Kingdom}\\
\vfill
\large {\bf Abstract}
\vspace{-0.2cm}
\end{center}
Second- and third-order results are presented for the structure functions of
charged-current deep-inelastic scattering in the framework of massless 
perturbative QCD. 
We write down the two-loop differences between the corresponding crossing-even 
and -odd coefficient functions, including those for the longitudinal structure
function not covered in the literature so far. At three loops we compute the
lowest five moments of these differences for all three structure functions and 
provide approximate expressions in Bjorken-$x$ space. 
Also calculated is the related third-order coefficient-function correction to 
the Gottfried sum rule. We confirm the conjectured suppression of these 
quantities if the number of colours is large.
Finally we derive the second- and third-order QCD contributions to the 
Paschos-Wolfenstein ratio used for the determination of the weak mixing angle 
from neutrino-nucleon deep-inelastic scattering. These contributions are found
to be small.
\vspace{0.5cm}
\end{titlepage}

\renewcommand{\theequation}{\thesection.\arabic{equation}}
%
%
\setcounter{equation}{0}
\section{Introduction}
\label{sec:introduction}
%
%
Structure functions in deep-inelastic scattering (DIS) are among the most 
extensively measured observables. Today the combined data from fixed-target 
experiments and the HERA collider spans about four orders of magnitude in both
Bjorken-$x$ and the scale $Q^2 = -q^2$ given by the momentum $q$ of the 
exchanged electroweak gauge boson \cite{Yao:2006px}. 
In this article we focus on the $W\!$-exchange charged-current (CC) case, see
Refs.~\cite {Yang:2000ju,Tzanov:2005kr,Onengut:2005kv} and 
\cite{Adloff:2003uh,Chekanov:2003vw,Aktas:2005ju,Chekanov:2006da} for recent 
measurements in neutrino DIS and at HERA. 
With six structure functions, $F_2^{\,W^\pm}\!$, $F_3^{\, W^\pm}$ and 
$F_L^{\,W^\pm}\!$, this case has a far richer structure than, for example, 
electromagnetic DIS with only two independent observables, $F_{\:\!2}$ 
and~$F_L$.
 
More detailed measurements are required to fully exploit the resulting 
potential, for instance at a future neutrino factory, see Ref.~\cite
{Mangano:2001mj}, and the LHeC, the proposed high-luminosity electron-proton
collider at the LHC~\cite{Dainton:2006wd}.
Already now, however, charged-current DIS provides important information on the
parton structure of the proton, e.g., its flavour decomposition and the 
valence-quark distributions. Moreover, present results are also sensitive to 
electroweak parameters of the Standard Model such as $\sin^2 \theta_W$, see 
Ref.~\cite{Zeller:2001hh}, and the space-like $W\!$-boson propagator
\cite{Aktas:2005iv}. As discussed, for example, in Refs.~\cite
{Davidson:2001ji,McFarland:2003jw,Dobrescu:2003ta,Kretzer:2003wy}, a reliable 
determination of $\sin^2 \theta_W$ from neutrino DIS requires a detailed 
understanding of non-perturbative and perturbative QCD effects.    

The perturbative calculations for the unpolarised structure functions in DIS 
have almost been completed to the next-to-next-to-leading order (NNLO) of 
massless QCD.  These results include the splitting functions, controlling the 
scale evolution of the parton distributions, to the third order in the strong 
coupling constant $\as$~\cite{Moch:2004pa,Vogt:2004mw}, as well as the 
hard-scattering coefficient functions for $F_1$, $F_{\:\!2}$ and $F_{\:\!3}$ to
second order in $\as$~\cite{SanchezGuillen:1991iq,vanNeerven:1991nn,%
Zijlstra:1991qc,Zijlstra:1992kj,Moch:1999eb}.
For the longitudinal structure function $F_L = F_{\:\!2} - 2x F_1$ the 
third-order coefficient functions are required at NNLO. So far these quantities
have been computed only for electromagnetic (photon-exchange) DIS
\cite{Moch:2004xu,Vermaseren:2005qc}. 
In fact, it appears that even the second-order coefficient functions for the 
charged-current $F_L$ have not been fully presented in the literature.  

It is convenient to consider linear combinations of the charged-current
structure functions $F_a^{\,W^\pm}$ with simple properties under crossing, such 
as $F_a^{\,\nu p \pm \bar \nu p}$ ($a = 2,\: 3,\: L$) for neutrino DIS. For all
these combinations either the even or odd moments can be calculated in 
Mellin-$N$ space in the framework of the operator product expansion (OPE), 
see Ref.~\cite{Buras:1980yt}.
The results for the third-order coefficient functions for the even-$N$ 
combinations $\, F_{2,L}^{\,\nu p + \bar\nu p}$ can be taken over from 
electromagnetic DIS \cite{Moch:2004xu,Vermaseren:2005qc}. Also the coefficient 
function for the odd-$N$ based quantity $\, F_3^{\,\nu p +\bar\nu p}$ is 
completely known at three-loop accuracy, with the results only published via 
compact parametrizations so far \cite{Vogt:2006bt}.
For the remaining combinations $\,F_{2,L}^{\,\nu p - \bar\nu p}$ and
$\,F_3^{\,\nu p - \bar\nu p\!}$, on the other hand, only the first five odd and 
even integer moments of the respective coefficient functions have been 
calculated to third order in Ref.~\cite{Moch:2007gx} following the 
approach of Refs.~\cite{Larin:1994vu,Larin:1997wd,Retey:2000nq} based on the
{\sc Mincer} program \cite{Gorishnii:1989gt,Larin:1991fz}.

The complete results of Refs.~\cite{Moch:2004xu,Vermaseren:2005qc,Vogt:2006bt} 
fix all even and odd moments $N$. Hence already the present knowledge is  
sufficient to determine also the lowest five moments of the differences of 
corresponding even-$N$ and odd-$N$ coefficient functions and to address a 
theoretical conjecture \cite{Broadhurst:2004jx} for these quantities.
Furthermore these moments facilitate $x$-space approximations in the style of, 
e.g, Ref.~\cite{vanNeerven:2001pe} which are sufficient for most 
phenomenological purposes, including the determination of the third-order QCD 
corrections to the Paschos-Wolfenstein relation~\cite{Paschos:1973kj} used for 
the extraction of $\sin^2 \theta_W$ from neutrino DIS.

The outline of this article is as follows.
In Section~\ref{sec:2-loop} we briefly specify our notations and write down the
complete second-order results $\delta\:\! c_{a}^{(2)}(x)$ for the above 
coefficient-function differences. We discuss their behaviour at the end points
$x = 0$ and $x = 1$, and provide compact but accurate parametrizations for use
in numerical applications. 
We then proceed, in Section~\ref{sec:3-loop}, to our new results for the five 
lowest odd moments of $\delta\:\! c_{2,L}^{(3)}$ and even moments of 
$\delta\:\! c_{3}^{(3)}\!$, as a byproduct deriving the third-order 
coefficient-function correction to the Gottfried sum rule. These three-loop 
moments are presented in a numerical form and employed to construct $x$-space 
approximations valid at $x \gsim 10^{-2}$. 
In Section~\ref{sec:applications} we address the numerical implications of 
our results. In particular we discuss the higher-order QCD corrections to the 
Paschos-Wolfenstein relation. 
Our~findings are finally summarized in Section~\ref{sec:summary}. The 
lengthy full expressions of the new third-order moments in terms of fractions 
and the Riemann \mbox{$\zeta$-function} can be found in the Appendix.
%
%
\setcounter{equation}{0}
\section{The complete second-order results}
\label{sec:2-loop}
%
%

We define the even-odd differences of the CC coefficient functions $\,C_a\,$ 
for $\,a = 2,\: 3,\: L\,$ as 
\bea
  \label{eq:cdiff}
  \delta\, C_{2,L} \; =\; C_{2,L}^{\,\nu p + {\bar \nu} p} 
    - C_{2,L}^{\,\nu p - {\bar \nu} p} \:\: , \qquad
  \delta\, C_3 \; =\;  C_3^{\,\nu p - {\bar \nu} p}
    - C_3^{\,\nu p + {\bar \nu} p} \:\: .
\eea
The signs are chosen such that the differences are always `even -- odd' in the 
moments $\, N$ accessible by the OPE \cite{Buras:1980yt}, and it is understood 
that the $d^{\:\!abc}d_{abc}$ part of $\,C_3^{\,\nu p + \bar\nu p}$ 
\cite{Retey:2000nq,Vogt:2006bt} is removed before the difference is formed. The
non-singlet quantities (\ref{eq:cdiff}) have an expansion in powers of $\as$, 
\bea
\label{eq:cf-exp}
  \delta\, C_a \; = \; 
  \sum_{l=2} \: \ar^{\, l}\: \delta\:\! c_{a}^{(l)} 
\eea
where, as throughout this and the next section, we are have normalized the 
expansion parameter as $\ar = \as /(4 \pi)$.
There are no first-order contributions to these differences, hence the sums
start at $l = 2\,$ in Eq.~(\ref{eq:cf-exp}).

All known DIS coefficient functions in massless perturbative QCD can be 
expressed in terms of the harmonic polylogarithms $H_{m_1,...,\,m_w}(x)$ with 
$m_j = 0,\,\pm 1$. Our notation for these functions follows Ref.~\cite
{Remiddi:1999ew} to which the reader is referred for a detailed discussion.
For $w \leq 3$  the harmonic polylogarithms can be expressed in terms of 
standard polylogarithms; a complete list can be found in Appendix A of 
Ref.~\cite{Moch:1999eb}. A {\sc Fortran} programs for these functions up to
weight $w=4$ has been provided in Ref.~\cite{Gehrmann:2001pz}, with an 
unpublished extension also covering $w=5$. In the remainder of this section
we employ the short-hand notation
\beq
\label{eq:habbr}
  H_{{\footnotesize \underbrace{0,\ldots ,0}_{\scriptstyle m} },\,
  \pm 1,\, {\footnotesize \underbrace{0,\ldots ,0}_{\scriptstyle n} },
  \, \pm 1,\, \ldots}(x) \; = \; H_{\pm (m+1),\,\pm (n+1),\, \ldots}(x)
\eeq
and additionally suppress the arguments of the harmonic polylogarithms for 
brevity. 

Exact expressions for (moments of) the coefficient functions will be given in 
terms of the SU($N_c$) colour factors $\,C_A = N_c\,$ and $\,C_F = (N_c^{\,2} 
-1)/(2N_c)$, while we use the QCD values $C_A = 3$ and $C_F = 4/3$ in numerical 
results. All our results are presented in the \MSb\ scheme for the standard
choice $\,\mu_r =\mu_f^{} = Q\,$ of the renormalization and factorization 
scales. 

The second-order coefficient functions $\delta\:\! c_2^{(2)}$ and 
$\,\delta c_L^{(2)}$ for the even-odd differences of $F_{\:\! 2,L\,}$ read
\bea
\label{eq:dc2qq2}
  \delta c_{2}^{(2)}(x) &\! =\! &
      \cf \* [\cf-\ca/2]  \*  \biggl(
          - {324 \over 5}
          + 112 \*  (1+x)^{-1} \* \z3
          + {16 \over 5} \* x^{-1}
          + {164 \over 5} \* x
          + {144 \over 5} \* x^2
  \nonumber\\ 
& &\mbox{}
         - 40 \* \z3
          + 136 \* \z3 \* x
          + 8 \* \z2
          + 56 \* \z2 \* x
          + 96 \* \z2 \* x^2
          - {144 \over 5} \* \z2 \* x^3
          - 32 \*  \Hh(-2,0)
  \nonumber\\
& &\mbox{}
          + 96 \*  \Hh(-2,0)  \*  (1+x)^{-1}
          + 128 \*  \Hh(-2,0)  \* x
          - 128 \*  \H(-1)  \*  (1+x)^{-1} \* \z2
          + 48 \*  \H(-1)  \* \z2
  \nonumber\\
& &\mbox{}
          - 144 \*  \H(-1)  \* \z2 \* x
          + 32 \*  \Hhh(-1,-1,0)
          - 128 \*  \Hhh(-1,-1,0)  \*  (1+x)^{-1}
          - 224 \*  \Hhh(-1,-1,0)  \* x
  \nonumber\\
& &\mbox{}
          + 64 \*  \Hh(-1,0)
          + {16 \over 5} \*  \Hh(-1,0)  \* x^{-2}
          + 64 \*  \Hh(-1,0)  \* x
          + 96 \*  \Hh(-1,0)  \* x^2
          - {144 \over 5} \*  \Hh(-1,0)  \* x^3
  \nonumber\\
& &\mbox{}
          - 64 \*  \Hhh(-1,0,0)
          + 160 \*  \Hhh(-1,0,0)  \*  (1+x)^{-1}
          + 160 \*  \Hhh(-1,0,0)  \* x
          + 64 \*  \Hh(-1,2)  \*  (1+x)^{-1}
  \nonumber\\
& &\mbox{}
          - 32 \*  \Hh(-1,2)
          + 32 \*  \Hh(-1,2)  \* x
          + {28 \over 5} \*  \H(0)
          - 32 \*  \H(0)  \*  (1+x)^{-1}
          - {16 \over 5} \*  \H(0)  \* x^{-1}
          - {292 \over 5} \*  \H(0)  \* x
  \nonumber\\
& &\mbox{}
          + 32 \*  \H(0)  \*  (1+x)^{-1} \* \z2
          + {144 \over 5} \*  \H(0)  \* x^2
          - 16 \*  \H(0)  \* \z2
          + 16 \*  \H(0)  \* \z2 \* x
          - 16 \*  \Hh(0,0)
          - 64 \*  \Hh(0,0)  \* x
  \nonumber\\
& &\mbox{}
          - 96 \*  \Hh(0,0)  \* x^2
          + {144 \over 5} \*  \Hh(0,0)  \* x^3
          + 24 \*  \Hhh(0,0,0)
          - 48 \*  \Hhh(0,0,0)  \*  (1+x)^{-1}
          - 24 \*  \Hhh(0,0,0)  \* x
  \nonumber\\
& &\mbox{}
          - 32 \*  \H(1)
          + 32 \*  \H(1)  \* x
          - 16 \*  \H(2)
          - 16 \*  \H(2)  \* x
          + 16 \*  \H(3)
          - 32 \*  \H(3)  \*  (1+x)^{-1}
          - 16 \*  \H(3)  \* x
         \biggr) 
\:\: ,
\\
\label{eq:dcLqq2}
  \delta c_{L}^{(2)}(x) &\! =\! &
      \cf \* [\cf-\ca/2]  \*  \biggl(
            {64 \over 5} \* x^{-1}
          - {416 \over 5}
          + {256 \over 5} \* x
          + {96 \over 5} \* x^2
          + 64 \* \z3 \* x
          + 32 \* \z2 \* x
          + 64 \* \z2 \* x^2
 \nonumber\\
& &\mbox{}
          - {96 \over 5} \* \z2 \* x^3
          + 64 \*  \Hh(-2,0)  \* x
          - 64 \*  \H(-1)  \* \z2 \* x
          - 128 \*  \Hhh(-1,-1,0)  \* x
          + 64 \*  \Hh(-1,0)
          + {64 \over 5} \*  \Hh(-1,0)  \* x^{-2}
 \nonumber\\
& &\mbox{}
          - 32 \*  \Hh(-1,0)  \* x^{-1}
          + 64 \*  \Hh(-1,0)  \* x
          + 64 \*  \Hh(-1,0)  \* x^2
          - {96 \over 5} \*  \Hh(-1,0)  \* x^3
          + 64 \*  \Hhh(-1,0,0)  \* x
          + {32 \over 5} \*  \H(0)
 \nonumber\\
& &\mbox{}
          - {64 \over 5} \*  \H(0)  \* x^{-1}
          - {448 \over 5} \*  \H(0)  \* x
          + {96 \over 5} \*  \H(0)  \* x^2
          - 32 \*  \Hh(0,0)  \* x
          - 64 \*  \Hh(0,0)  \* x^2
          + {96 \over 5} \*  \Hh(0,0)  \* x^3
          \biggr)
\:\: .  \quad 
\eea
The corresponding quantity $\delta\:\! c_3^{(2)}$ for the charged-current 
structure functions $F_{\:\! 3}$ is given by
\bea
\label{eq:dc3qq2}
  \delta c_{3}^{(2)}(x) &\! =\! & \delta c_{2}^{(2)}(x) \;
   - \cf \* [\cf-\ca/2]  \*  \biggl(
          - {624\over 5}
          + {16\over 5} \* x^{-1}
          + {464\over 5} \* x
          + {144\over 5} \* x^2
          + 32 \* \z3
\nonumber \\
& &\mbox{}
          + 96 \* \z3 \* x 
          - 16 \* \z2
          + 48 \* \z2 \* x
          + 80 \* \z2 \* x^2
          - {144\over 5} \* \z2 \* x^3
          + 32 \* \Hh(-2,0)
          + 96 \* \Hh(-2,0) \* x
\nonumber\\
& &\mbox{}
          - 32 \* \H(-1) \* \z2
          - 96 \* \H(-1) \* \z2 \* x
          - 64 \* \Hhh(-1,-1,0)
          - 192 \* \Hhh(-1,-1,0) \* x
          + 64 \* \Hh(-1,0)
\nonumber\\
& &\mbox{}
          + {16\over 5} \* \Hh(-1,0) \* x^{-2}
          - 16 \* \Hh(-1,0) \* x^{-1}
          + 64 \* \Hh(-1,0) \* x
          + 80 \* \Hh(-1,0) \* x^2
          - {144\over 5} \* \Hh(-1,0) \* x^3
\nonumber\\
& &\mbox{}
          + 32 \* \Hh(-1,0,0)
          + 96 \* \Hh(-1,0,0) \* x
          - {16\over 5} \* \H(0) \* x^{-1}
          - {112\over5} \* \H(0)
          - {592\over 5} \* \H(0) \* x
          + {144\over 5} \* \H(0) \* x^2
\nonumber\\
& &\mbox{}
          + 16 \* \Hh(0,0)
          - 48 \* \Hh(0,0) \* x
          - 80 \* \Hh(0,0) \* x^2
          + {144\over 5} \* \Hh(0,0) \* x^3
          \biggr)
\:\: . 
\eea
Expressions equivalent to Eqs.~(\ref{eq:dc2qq2}) and (\ref{eq:dc3qq2}) have
first been published in Refs.~\cite{vanNeerven:1991nn} and \cite
{Zijlstra:1992kj}, respectively, and were later confirmed in Ref.~\cite
{Moch:1999eb}. To the best of our knowledge, on the other hand, the function 
$\,\delta c_L^{(2)}$ has not been documented in the literature before, see, 
e.g., Ref.~\cite{Kazakov:1990fu} and references therein. 
It was however calculated by the authors of Refs.~\cite
{vanNeerven:1991nn,Zijlstra:1991qc,Zijlstra:1992kj}, distributed in a 
{\sc Fortran} package of the two-loop coefficient functions, and employed for
the parametrizations of Ref.~\cite{vanNeerven:1999ca}. Our expression 
(\ref{eq:dcLqq2}) agrees with this unpublished result.

It is instructive to briefly consider the end-point limits of the above 
results. Suppressing the ubiquitous factor $\,C_F C_{FA} \equiv 
C_F [\,C_F-C_A/2]$, the small-$x$ behaviour of Eqs.~(\ref{eq:dc2qq2}) -- 
(\ref{eq:dc3qq2}) is 
\bea
\label{eq:c2-smallx}
 \delta c_{2}^{(2)}(x) &\! \simeq\! & \mbox{}
   - 4\, \ln^3 x \:  - \;\: 8\, \ln^2 x - ( 28 - 16\,\z2 ) \ln x 
   \: - \: 64 \: + \;\: 8\,\z2 + 72\,\z3 \: + \: \ldots
 \nn \\
 \delta c_{3}^{(2)}(x) &\! \simeq\! & \mbox{}
   - 4\, \ln^3 x - 16\, \ln^2 x + ( 12 + 16\,\z2 ) \ln x 
   \: + \: 44 + 24\,\z2 + 40\,\z3 \: + \: \ldots
 \nn \\[1mm]
 \delta c_{L}^{(2)}(x) &\! \simeq\! & \mbox{}
   - 32\, \ln x - 48 \: + \: \ldots \:\: .
\eea
Thus the even-odd differences are not suppressed with respect to the $\,\nu p 
+ \bar\nu p\,$ two-loop non-singlet coefficient functions for $x\ra 0\, $: 
the same powers of $\,\ln x\,$ enter Eqs.~(\ref{eq:c2-smallx}) and those 
quantities. At large $x$, on the other hand, all three functions 
$\delta\:\! c_{a}^{(2)}$ are suppressed by factors $(1-x)^2$ times logarithms, 
reading
\bea
\label{eq:c2-largex}
 \delta c_{2}^{(2)}(x) &\! = \! & - ( 12 - 8\,\z2 )\, [1-x] \: C_F C_{FA}
   \: + \: O \left( [1-x]^2 \right)
 \nn \\
 \delta c_{3}^{(2)}(x) &\! = \! & \phantom- ( 20 - 8\,\z2 )\, [1-x] \: 
   C_F C_{FA} \: + \: O \left( [1-x]^2 \right)
 \nn \\
 \delta c_{L}^{(2)}(x) &\! = \! & ( 32 -16\,\z2 )\, [1-x]^2 \: C_F C_{FA}
   \: + \: O \left( [1-x]^3 \right) \:\: .
\eea
  
\begin{figure}[th]
\centerline{\epsfig{file=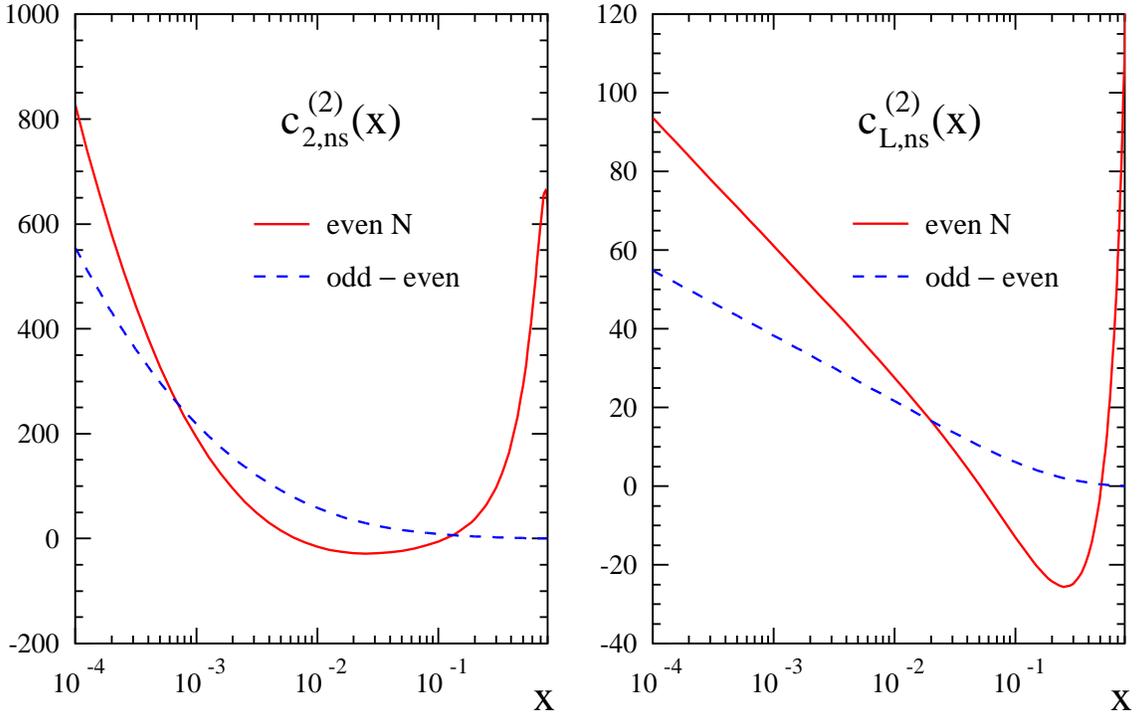,width=15cm,angle=0}}
\vspace{-2mm}
\caption{The odd$\,-\,$even non-singlet differences $\, -\, \delta\:\! 
 c_{2,L}^{(2)}(x)$ of Eqs.~(\ref{eq:dc2qq2}) and (\ref{eq:dcLqq2}), compared at
 $x\leq 0.8$ to the corresponding even-$N$ coefficient functions calculated in 
 Refs.~\cite {SanchezGuillen:1991iq,vanNeerven:1991nn,Moch:1999eb}.
 \label{fig:c2diff} } 
\vspace*{1mm}
\end{figure}

\vspace{-2mm}
The differences $\delta\:\! c_{2}^{(2)}(x)$ and $\delta\:\! c_{L}^{(2)}(x)$ 
(both multiplied by -1 for display purposes) are compared to the corresponding 
even-$N$ $\,\nu p +\bar\nu p\,$ coefficient functions in Fig.~\ref{fig:c2diff}.
The quantities (\ref{eq:dc2qq2}) and (\ref{eq:dcLqq2}) are negligible at 
$\,x \gsim 0.1\,$ and at $\,x \gsim 0.3\,$, respectively, but indeed comparable
to the even-moment coefficient functions at small $x$. The corresponding 
results for $F_{\:\!3}$ are qualitative similar to those for $F_{\:\!2}$, but 
with $\delta\:\! c_{3}^{(2)}(x)$ small down to $\,x \simeq 0.01\,$.

For certain numerical applications, for instance for use with complex-$N$
packages like Ref.~\cite{Vogt:2004ns}, it is convenient to have 
parametrizations of Eqs.~(\ref{eq:dc2qq2}) -- (\ref{eq:dc3qq2}) in terms of 
elementary functions. With an error of less than 0.1\% these functions can be
approximated by
\bea
\label{eq:dc2qq2p}
 \delta c_{2}^{(2)}(x) &\! \simeq\! & 
 \{ - 9.1587 - 57.70\, x + 72.29\, x^2 - 5.689\, x^3
  - xL_0\, ( \, 68.804 + 24.40\, L_0 
 \nn \\ & & \mbox{}
  \; + 2.958\, L_0^2 \: )
  + 0.249\, L_0 + 8/9\: L_0^2\, (2 + L_0) \, \} \: (1 - x)
 \:\: , \nn \\
 \delta c_{3}^{(2)}(x) &\! \simeq\! &
  \{ - 29.65 + 116.05\, x - 71.74\, x^2 - 16.18\, x^3
  + xL_0\, ( \, 14.60 + 69.90\, x 
 \nn \\ & & \mbox{}
 - 0.378\, L_0^2 \: ) 
 - 8.560\, L_0 + 8/9\: L_0^2\, (4 + L_0) \, \} \: (1 - x) 
 \:\: , \nn \\
 \delta c_{L}^{(2)}(x) &\! \simeq\! &
  \{ \, 10.663 - 5.248\, x - 7.500\, x^2 + 0.823\, x^3
  + xL_0\, ( \, 11.10 + 2.225\, L_0
 \nn \\ & & \mbox{}
  \; - 0.128\, L_0^2 \: ) 
  + 64/9\: L_0 \, \} \: (1 - x)^2
 \:\: .
\eea
Here we have employed the short-hand $\, L_0 = \ln x\,$ and inserted the QCD
values of $C_F$ and $C_A$.
%
%
\setcounter{equation}{0}
\section{Third-order moments and approximations}
\label{sec:3-loop}
%
%
Recently the first five odd-integer moments have been computed of the third-%
order coefficient functions for $\,F_{2,L}^{\,\nu p - \bar\nu p}$ in 
charged-current DIS, together with the corresponding moments 
$\,N = 2,\, \ldots ,\, 10\,$ for $\,F_3^{\,\nu p - \bar\nu p\!}$ 
\cite{Moch:2007gx}. Unlike previous fixed-$N$ calculations, the complete
three-loop results for $F_{2,L}^{\,\nu p + \bar \nu p\,}$ 
\cite{Moch:2004xu,Vermaseren:2005qc}\footnote
{$\,$The $\as^3$ coefficient functions for this process are those of 
photon-exchange DIS, but without the contributions of the $fl_{11}$ flavour
classes, see Fig.~1 of Ref.~\cite{Vermaseren:2005qc}, where the two photons 
couple to different quark loops.} 
and $F_3^{\nu P + \bar \nu P}$ \cite{Vogt:2006bt} facilitate analytic 
continuations to these values of $N$. We have performed this continuation 
using the $x$-space expressions in terms of harmonic polylogarithms \cite
{Remiddi:1999ew} and the Mellin transformation package provided with version 3 
of {\sc Form}~\cite{Vermaseren:2000nd}. 
Thus we are in a position to derive the respective lowest five moments of
the hitherto unknown third-order contributions to the even-odd differences
(\ref{eq:cdiff}). These moments represent the main new results of this article.
With one exception (see below) the exact SU($N_c$) expressions are however 
deferred to the Appendix.

Here we present numerical results for QCD, using the conventions introduced at 
the beginning of Section~\ref{sec:2-loop}, recall especially $\,\ar\, \equiv\, 
\as/(4 \pi)\,$ and the scale choice $\,\mu_r = \mu_f^{} = Q\,$.
In addition $\nf$ denotes the number of effectively massless quark flavours, 
and we use the notation $\delta\, C_{a,\,N}$ for the $N$-th moment of
$\delta\, C_{a}(x)$. The results for $F_{\:\!2}$ and $F_L$ read 
\bea
\label{eq:dc2ns}
  \delta\, C_{2,1} & = &
          - 4.378539253
           \, \ar^2
       \: + \:\ar^3 \,  \*  (
          - 125.2948456
          - 0.6502282123 \* \, \nf
          )
\nonumber
\\
  \delta\, C_{2,3} & = &
          - 0.138066958
          \, \ar^2
       \: + \:\ar^3 \,  \*  (
          - 5.554493975
          + 0.1939792023 \* \, \nf
          )
\nonumber
\\
  \delta\, C_{2,5} & = &
          - 0.032987989
           \, \ar^2
       \: + \:\ar^3 \,  \*  (
          - 0.707322026
          + 0.0004910378 \* \, \nf
          )
\nonumber
\\
  \delta\, C_{2,7} & = &
          - 0.013235254
           \, \ar^2
       \: + \:\ar^3 \,  \*  (
          - 0.008816536
          - 0.0201069660 \* \, \nf
          )
\nonumber
\\
  \delta\, C_{2,9} & = &
          - 0.006828983
           \, \ar^2
       \: + \:\ar^3 \,  \*  (
           \phantom+ 0.133159220
          - 0.0200289710 \* \, \nf
          )
\eea
and
\bea
\label{eq:dcLns}
  \delta\, C_{L,1} & = &
          - 2.138954096
           \, \ar^2
       \: + \:\ar^3 \,  \*  (
          - 106.6667685
          + 3.294301343 \* \, \nf
          )
\nonumber
\\
  \delta\, C_{L,3} & = &
          - 0.078259985
          \, \ar^2
       \: + \:\ar^3 \,  \*  (
          - 9.239637919
          + 0.2718024935 \* \, \nf
          )
\nonumber
\\
  \delta\, C_{L,5} & = &
          - 0.016892540
           \, \ar^2
       \: + \:\ar^3 \,  \*  (
          - 2.548566852
          + 0.0650677125 \* \, \nf
          )
\nonumber
\\
  \delta\, C_{L,7} & = &
          - 0.006263113
           \, \ar^2
       \: + \:\ar^3 \,  \*  (
          - 1.075400460
          + 0.0251053847 \* \, \nf
          )
\nonumber
\\
  \delta\, C_{L,9} & = &
          - 0.003001231
           \, \ar^2
       \: + \:\ar^3 \,  \*  (
          - 0.560603262
          + 0.0122952192 \* \, \nf
          )
 \:\: .
\eea
The lowest even moments for the structure function $F_{\:\! 3}$ are given by
\bea
\label{eq:dc3ns}
  \delta\, C_{3,2} & = &
          - 0.1135841071
           \, \ar^2
       \: + \:\ar^3 \,  \*  (
          \phantom+ 8.386266870
          + 0.0605431788 \* \, \nf
          )
\nonumber
\\
  \delta\, C_{3,4} & = &
          - 0.0683669250
           \, \ar^2
       \: + \:\ar^3 \,  \*  (
          - 1.237248886
          + 0.0971522112  \* \, \nf
          )
\nonumber
\\
  \delta\, C_{3,6} & = &
          - 0.0350849853
          \, \ar^2
       \: + \:\ar^3 \,  \*  (
          - 1.370404531
          + 0.0496762716 \* \, \nf
          )
\nonumber
\\
  \delta\, C_{3,8} & = &
          - 0.0208455457
          \, \ar^2
       \: + \:\ar^3 \,  \*  (
          - 1.052847874
          + 0.0282541123 \* \, \nf
          )
\nonumber
\\
  \delta\, C_{3,10} \!\! & = &
          - 0.0137316528
          \, \ar^2
       \: + \:\ar^3 \,  \*  (
          - 0.798850682
          + 0.0177100327 \* \, \nf
          )
 \:\: .
\eea
The new $\as^3$ contributions are rather large if compared to the leading
second-order results also included in Eqs.~(\ref{eq:dc2ns}) -- (\ref{eq:dc3ns})
with, e.g., $\,\ar = 1/50\,$ corresponding to $\,\as \simeq 0.25$. Except for 
the lowest moment for $\,a = 2,L$, on the other hand, the integer-$N$ 
differences $\delta\, C_{a,N}$ are entirely negligible compared to the 
$\,\nu p \pm \bar\nu p\,$ moments of Refs.~\cite{Retey:2000nq,Moch:2007gx}.
 
Before we turn to the $x$-space implications of Eqs.~(\ref{eq:dc2ns}) -- 
(\ref{eq:dc3ns}), let us briefly discuss some interesting structural features
of our third-order results. For this purpose we consider the exact SU($N_c$) 
expression for the lowest moment of $\,\delta\:\! c_2^{(3)}$ given by
\bea
\label{eq:dc2q1}
  \delta c_{2,1}^{(3)} & = &
       \cf \* \cfas \* \Biggl(
          {175030 \over 81}
          - {49216 \over 27} \* \z2
          + {404720 \over 81}\*\z3
          - {562784 \over 135} \* \zs2
          + {33200 \over 9}\* \z2 \* \z3
\nonumber\\
& &\mbox{}\qquad\qquad
          - {4160 \over 9} \*\z5
          - {8992 \over 63} \* \zt2
          - {1472 \over 3}\*\zs3
          \Biggr)
\nonumber\\[0.5mm]
& &\mbox{\hspn}
       + \cfs \* \cfa \* \Biggl(
          -{ 303377 \over 162}
          +{ 41350 \over 27} \* \z2
          -{ 363896 \over 81} \* \z3
          +{ 396824 \over 135} \* \zs2
          -{ 26000 \over 9} \* \z2 \* \z3
\nonumber\\
& &\mbox{}\qquad\qquad
          +{ 25616 \over 9} \* \z5
          +{ 1456 \over 3}  \* \zs3
          -{ 56432 \over 315} \*\ \zt2
          \Biggl)
\\[0.5mm]
& &\mbox{\hspn}
       + \cf \* \cfa \* \nf \* \Biggl(
          { 8786 \over 81}
          - {3056 \over 27} \* \z2
          + {39592 \over 81} \* \z3
          +{ 1408 \over 9} \* \z2 \* \z3
          -{ 30424 \over 135} \* \zs2
          -{ 1792 \over 9} \* \z5
          \Biggr)
\nonumber 
\:\: .\quad 
\eea
As all other calculated moments of the functions $\delta\:\! c_{a}^{(2)}(x)$,
this result contains an overall factor $\cfa\,=\, C_F - C_A /2\,=\, -1/(2N_c)$.
Hence the third-order even-odd differences are suppressed in the large-$N_c$ 
limit as conjectured, to all orders, in Ref.~\cite{Broadhurst:2004jx} on the 
basis of two-loop results in particular for $N=1$ Adler and Gottfried sum 
rules, for a recent discussion see also Ref.~\cite{Kataev:2007jz}. 
In fact, up to the additional $fl_{11}$ contribution absent in charged-current 
DIS (recall Footnote 1),
\bea
\label{dc2em}
  \Delta_{\:\rm e.m.}\, c_{2,1}^{(3)} &\!= \! & 
  \dabcnc \, \biggl( - 288 
            + 96\,\z2 
            + {1472 \over 3}\, \z3 
            - {256 \over 5}\, \zs2 
            - {1280 \over 3}\,\z5 \biggr)
  \nn \\[1mm]
  &\!= \! & - 33.67693293\:\nf \qquad \mbox{in~~QCD} \:\: ,
\eea
Eq.~(\ref{eq:dc2q1}) represents the third-order coefficient-function 
correction to the Gottfried sum rule (GSR)\footnote
{$\,$Note that our overall normalization and expansion parameter differ from 
those of Ref.~\cite{Broadhurst:2004jx}. Consequently the corresponding GSR
coefficients (\ref{eq:dc2ns}), (\ref{eq:dc2q1}) and (\ref{dc2em}) are 
larger by a factor $4^l/3$ at order $\as^{\: l}$ than in their notation.}$\!\!$,
since the Adler sum rule involving the non-singlet coefficient function 
$C_{2,1}$ of the $\,\nu p - \bar\nu p \,$ combination does not receive any 
perturbative or non-perturbative corrections, see, e.g., Ref.~\cite
{Dokshitzer:1995qm}.

Another interesting feature of the functions $\delta\:\! c_{a=2,3}^{(l)}$ in 
Eq.~(\ref{eq:cf-exp}) is the presence of $\zeta$-functions up to weight $2l\,$ 
in the integer moments, e.g., terms up to $\zt2$ and $\zs3$ occur in the 
third-order result (\ref{eq:dc2q1}). This is in contrast to the `natural'
(OPE-based) moments of $C_a^{\nu p \pm \bar\nu p}$ which only include 
contributions up to weight $2l\! -\! 1$, see 
Refs.~\cite{Larin:1994vu,Larin:1997wd,Retey:2000nq,Moch:2007gx}.
Yet the $x$-space expressions of all these quantities consist of harmonic
polylogarithms up to weight $2l\! -\! 1$ corresponding to harmonic sums up to
weight $2l$. Note also that, in the approach of Refs.~\cite{vanNeerven:1991nn,%
Zijlstra:1991qc,Zijlstra:1992kj}, the absence of weight-$2l\,$ terms in the 
natural moments appears to require a cancellation between different diagram 
classes.

We now return to the numerical moments (\ref{eq:dc2ns}) -- (\ref{eq:dc3ns})
and investigate their consequences for the $x$-space functions 
$\delta\:\! c_{a}^{(3)}(x)$. We follow an approach successfully used, for 
instance, in Ref.~\cite{vanNeerven:2001pe} when only the coefficient-functions 
moments of Refs.~\cite{Larin:1994vu,Larin:1997wd,Retey:2000nq} were known. 
Based on the two-loop end-point behaviour in Eqs.~(\ref{eq:c2-smallx}) and 
(\ref{eq:c2-largex}) we expect small-$x$ terms up to $\ln^5 x$ and $\ln^3 x$ 
in $\,\delta\:\! c_{2,3}^{(3)}(x)\,$ and $\,\delta\:\! c_{L}^{(3)}(x)$, 
respectively, and large-$x$ limits including contributions up to 
$\,(1-x)^{\eta_a} \ln^2 (1-x)\,$ with $\,\eta_{2,3}^{} = 1\,$ and 
$\,\eta_{L}^{} = 2$. 
Thus the $x$-space expressions of $\delta\:\! c_{a}^{(3)}$ will be of the form
\beq 
  \delta c_{a}^{(3)}(x) \; = \; (1-x)^{\eta_a}\: \bigg( 
  \sum_{m=1}^2 A_{m}\,\ln^{\,m}(1-x) \, + \, \delta c_{a}^{\:\rm smooth}(x) 
  \, + \, B_1\, \frac{\ln x}{1-x} \bigg) 
  \: + \! \sum_{n=2}^{\;\;7-2\eta_a} B_n\, \ln^{\,n} x \quad
\eeq
where the functions $\delta\:\! c_{a}^{\:\rm smooth}(x)$ are finite for 
$0 \leq  x \leq 1$. For moment-based approximations a simple ansatz is chosen 
for these functions, and its free parameters are determined from the available 
moments together with a reasonably balanced subset of the coefficients $A_m$ 
and $B_n$. This ansatz and the choice of the non-vanishing end-point parameters
are then varied in order to estimate the remaining uncertainties of $\delta\:\! 
c_{a}^{(3)}(x)$. Finally for each value of $a$ two (out of about 50) 
approximations, denoted below by $A$ and $B$, are selected which indicate the
widths of the uncertainty bands. 
 
\begin{figure}[bh]
\vspace{-3mm}
\centerline{\epsfig{file=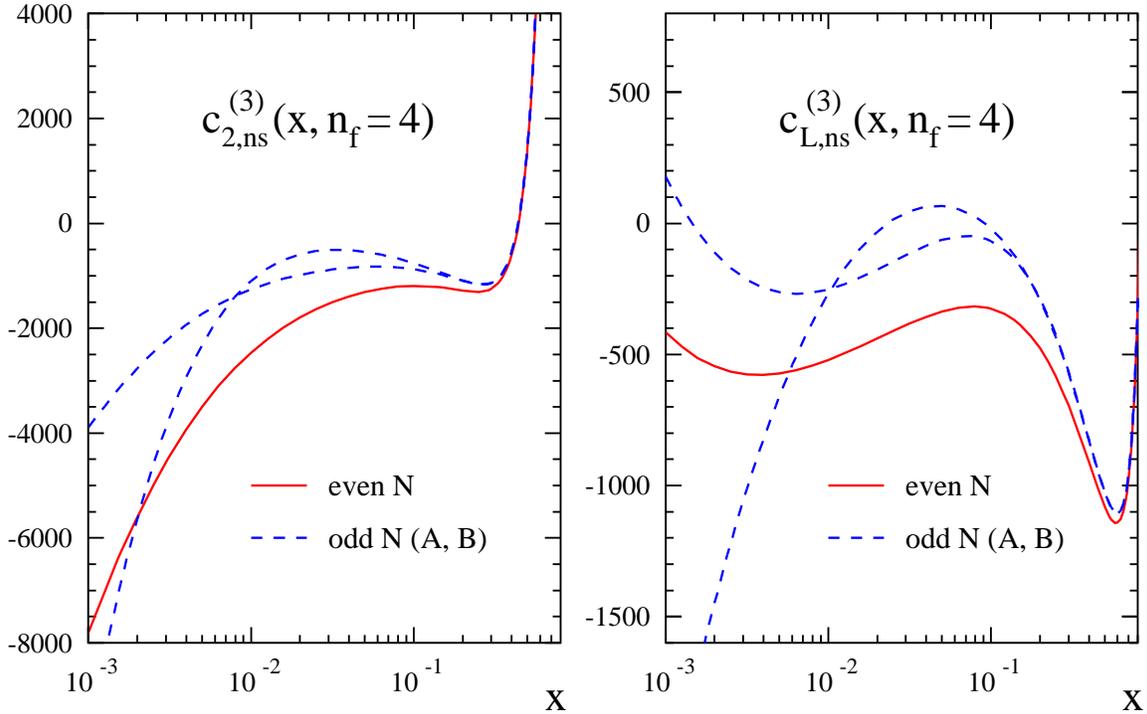,width=15cm,angle=0}}
\vspace{-2mm}
\caption{The exact third-order coefficient functions of the even-$N$ structure 
 functions $\, F_{2,L}^{\,\nu p + \bar\nu p}$ 
 \cite{Moch:2004xu,Vermaseren:2005qc} for four massless flavours, and the 
 corresponding odd-moment quantities obtained from these results and the 
 approximations (\ref{eq:dc2qq3p}) and (\ref{eq:dcLqq3p}) for the even -- odd 
 differences. \label{fig:c3diff}}
\vspace*{-4mm}
\end{figure}

For $F_{\:\! 2}$ and $F_L$ these functions are, with $\, L_0 = \ln x\,$, 
$x_1 = 1-x$  and $\, L_1 = \ln x_1$, 
\bea
\label{eq:dc2qq3p}
 \delta c_{2,\,A}^{(3)}(x) &\! =\! &
   ( 54.478\,L_1^2 + 304.6\,L_1 + 691.68\, x ) \, x_1
   + 179.14\,L_0 - 0.1826\,L_0^3
 \nn \\ & & \mbox{\hspn}
 + \nf\, \{ ( 20.822\, x^2 - 282.1\, (1 + {\textstyle {x \over 2}}) )\, x_1
   - (285.58\, x + 112.3 - 3.587\,L_0^2) \, L_0 \}
 \:\: , \nn \\[0.5mm]
 \delta c_{2,\,B}^{(3)}(x) &\! =\! &
   - ( 13.378\,L_1^2 + 97.60\,L_1 + 118.12\, x ) \, x_1
   - 91.196\,L_0^2 - 0.4644\,L_0^5  
 \\ & & \mbox{\hspn}
 + \nf\, \{ (4.522\,L_1 + 447.88\, (1 + {\textstyle {x \over 2}}) ) \, x_1
   + (514.02\, x + 147.05 + 7.386\,L_0)\, L_0 \} 
 \quad \nn 
\eea
and
\bea
\label{eq:dcLqq3p}
 \delta c_{L,\,A}^{(3)}(x) &\! =\! &
   - ( 495.49\,x^2 + 906.86 ) \, x_1^{\,2} - 983.23\,x x_1 L_0
   + 53.706\,L_0^2 + 5.3059\,L_0^3
 \nn \\ & & \mbox{\hspn}
 + \nf\, \{ ( 29.95\, x^3 - 59.087\,x^2 + 379.91 ) \, x_1^{\,2}
   - 273.042\, xL_0^2 + 71.482\, x_1L_0 \}
 \:\: , \nn \\[0.5mm]
 \delta c_{L,\,B}^{(3)}(x) &\! =\! &
   ( 78.306\,L_1 + 6.3838\, x ) \, x_1^{\,2} + 20.809\,x x_1 L_0
   - 114.47\,L_0^2 - 22.222\,L_0^3
 \\ & & \mbox{\hspn}
 + \nf\, \{ (12.532\,L_1 + 141.99\,x^2 - 250.62\, x ) \, x_1^{\,2}
   - ( 153.586\,x - 0.6569 ) \, x_1L_0 \}
 \:\: . \quad \nn
\eea
The corresponding results for $F_{\:\! 3}$ read
\bea
\label{eq:dc3qq3p}
 \delta c_{3,\,A}^{(3)}(x) &\! =\! &
   ( 3.216\,L_1^2 + 44.50\,L_1 - 34.588 ) \, x_1
   + 98.719\,L_0^2 + 2.6208\,L_0^5
 \nn \\ & & \mbox{\hspn}
 - \nf\, \{ ( 0.186\, L_1 + 61.102\, (1 + x) ) \, x_1
   + 122.51\, xL_0 - 10.914\,L_0^2 - 2.748\,L_0^3 \}
 \:\: , \nn \\[0.5mm]
 \delta c_{3,\,B}^{(3)}(x) &\! =\! &
   - ( 46.72\,L_1^2 + 267.26\,L_1 + 719.49\, x ) \, x_1
   - 171.98\,L_0 + 9.470\,L_0^3
 \\ & & \mbox{\hspn}
 + \nf\, \{ (0.8489\,L_1 + 67.928\, (1 + {\textstyle {x \over 2}}) ) \, x_1
   + 97.922\, xL_0 - 17.070\,L_0^2 - 3.132\,L_0^3 \} 
 \:\: . \quad \nn
\eea
The resulting approximations for the $\,\nu p -\bar\nu p\,$ odd-$N$ coefficient
functions $c_{2,L}^{(3)}(x)$ are compared in Fig.~\ref{fig:c3diff} to their
exact counterparts \cite{Moch:2004xu,Vermaseren:2005qc} for the even-$N$
non-singlet structure functions. The third-order even-odd differences remain
noticeable to larger values of $x$ than at two loops, e.g., up to $x \simeq
0.3$ for $F_{\:\!2}$ and $x \simeq 0.6$ for $F_{\:\! L}$ for the four-flavour
case shown in the figure. The moments $N = 1,\:3,\:\ldots,\: 9\,$ constrain
$\,\delta\:\! c_{2,L}^{(3)}(x)\,$ very well at $\,x \gsim 0.1$, and
approximately down to $\,x \approx 10^{-2}$.
 
For some applications, such as the Paschos-Wolfenstein relation addressed in the
next section, one needs the second moments of the functions $\,\delta\:\! 
c_{2,L}^{(3)}(x)$. These quantities can now be determined approximately from 
the above $x$-space results, yielding  
\bea
\label{eq:dc3mom2}
  \delta c_{2,2}^{(3)} &\! =\! & -20.19 \pm 0.39 \: + \: (0.691\pm 0.040) \,\nf
\nn \\
  \delta c_{L,2}^{(3)} &\! =\! & -24.75 \pm 0.15 \: - \: (0.792\pm 0.014) \,\nf
\:\: .
\eea
Here the central values are given by the respective averages of the 
approximations $A$ and $B$ in Eqs.~(\ref{eq:dc2qq3p}) and (\ref{eq:dcLqq3p})
which directly provide the upper and lower limits.

\begin{figure}[th]
\centerline{\epsfig{file=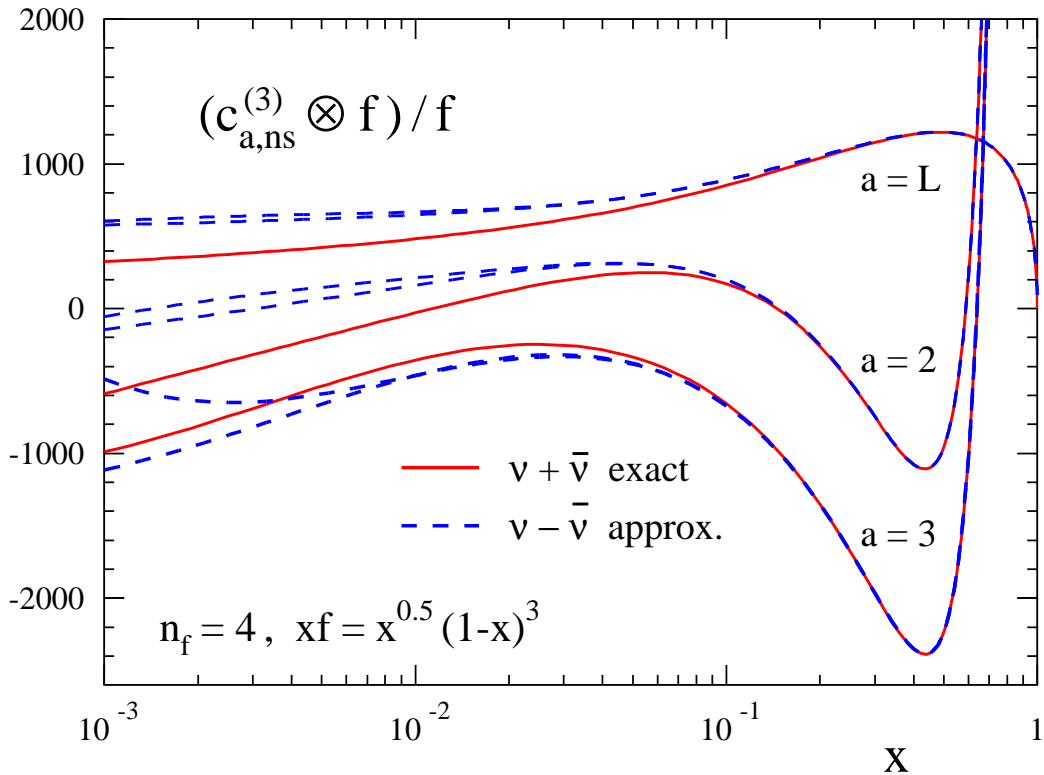,width=14cm,angle=0}\qquad}
\vspace*{-1mm}
\caption{Convolution of the six third-order CC coefficient functions for 
 $F_{\:\!2,\,3,\,L}$ in $\,\nu p + \bar\nu p\,$ 
 \cite{Moch:2004xu,Vermaseren:2005qc,Vogt:2006bt} and $\,\nu p - \bar\nu p\,$
 [Eqs.~(\ref{eq:dc2qq3p}) -- (\ref{eq:dc3qq3p})] DIS with a schematic but 
 typical non-singlet distribution $\!f$. All results have been normalized to 
 $\!f(x)$, suppressing a large but trivial variation of the absolute 
 convolutions for small and large values of $x$.  \label{fig:c3dcnv} }
\vspace*{-2mm}
\end{figure}

Returning to $x$-space we recall that 
uncertainty bands as in Fig.~\ref{fig:c3diff} do not directly indicate the
range of applicability of these approximations, since the coefficient functions
enter observables only via smoothening Mellin convolutions with
non-perturbative initial distributions. In Fig.~\ref{fig:c3dcnv} we therefore
present the convolutions of all six third-order CC coefficient functions with
a characteristic reference distribution. It turns out that the approximations
(\ref{eq:dc2qq3p}) and (\ref{eq:dcLqq3p}) of the previous figure can be
sufficient down to values even below $x = 10^{-3}$. The uncertainty of
$\,\delta\:\! c_{3}^{(3)}(x)$, on the other hand, becomes relevant already
at larger values, $\,x \lsim 10^{-2}$, as the lowest calculated moment of this
quantity, $\,N=2$, has far less sensitivity to the behaviour at low $x$.

The three-loop corrections to the non-singlet structure functions are rather
small even well below the $x$-values shown in the figure~~--~~recall our small 
expansion parameter $\ar\,$: the third-order coefficient are smaller by a 
factor $2.0\cdot 10^{-3}$ if the expansion is written in powers of $\as$. 
Their sharp rise for $\,x \ra 1\,$ is understood in terms of soft-gluon effects
which can be effectively resummed, if required, to next-to-next-to-next-to-%
leading logarithmic accuracy \cite{Moch:2005ba}. Our even-odd differences 
$\,\delta\:\! c_{a}^{(3)}(x)$, on the other hand, are irrelevant at $x > 0.1$ 
but have a sizeable impact at smaller $x$ in particular on the corrections for 
$F_{\:\!2}$ and $F_{\:\!L}$.
%
%
\setcounter{equation}{0}
\section{Applications}
\label{sec:applications}
\begin{figure}[hbt]
\centerline{\epsfig{file=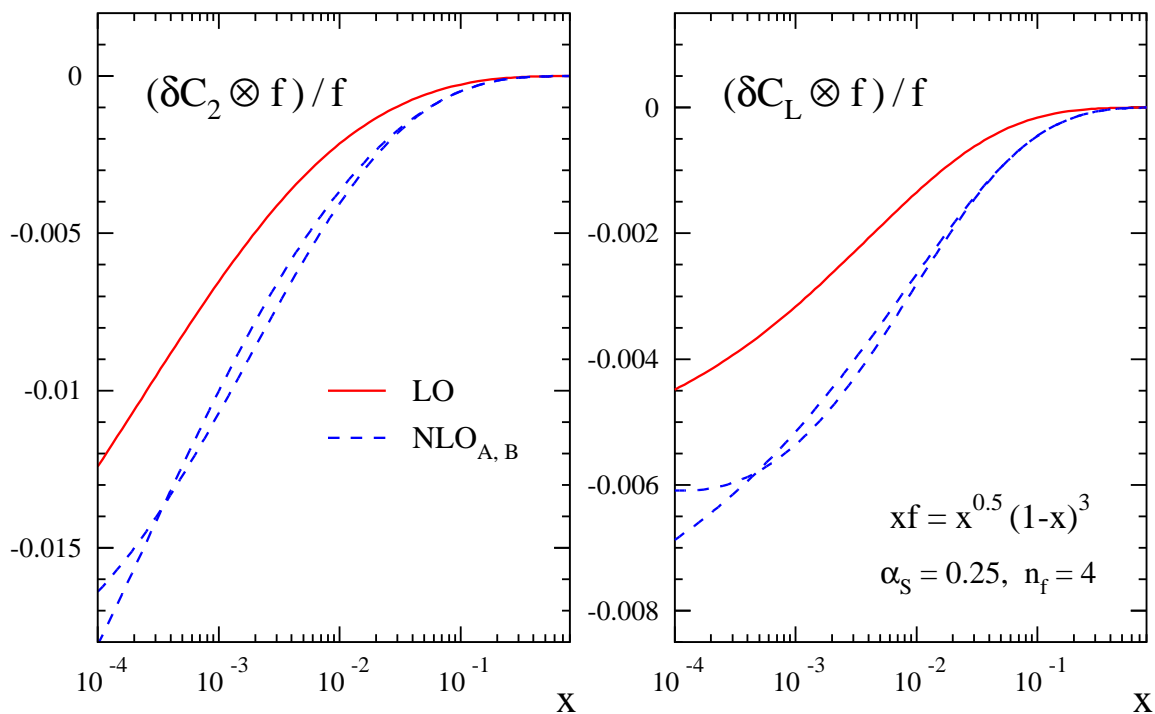,width=15.2cm,angle=0}\hspace{3mm}}
\vspace{-1mm}
\caption{The first two approximations, denoted by LO and NLO, of the 
 differences (\ref{eq:cf-exp}) for $F_{\:\!2}$ and $F_{\:\!L}$ in 
 charged-current DIS. The results are shown for representative values of 
 $\as$ and $\nf$ after convolution with the reference distribution $\! f(x)$
 also employed in Fig.~3. The dashed curves correspond to the two 
 approximations in Eqs.~(\ref{eq:dc2qq3p}) and (\ref{eq:dcLqq3p}) for the new 
 $\as^{\,3}$ contributions. \label{fig:c2lexp} }
\vspace*{-2mm}
\end{figure}

%
%
The approximate results for $\,\delta\:\! c_{a}^{(3)}(x)$ facilitate a first 
assessment of the perturbative stability of the even-odd differences 
(\ref{eq:cdiff}). In Fig.~\ref{fig:c2lexp} we illustrate the known two orders 
for $F_{\:\!2}$ and $F_{\:\!L}$ for \mbox{$\as = 0.25$} and $\nf = 4$ massless 
quark flavours, employing the same reference quark distribution as in 
Fig.~\ref{fig:c3dcnv}. 
Obviously our new $\as^{\,3}$ corrections are important wherever these 
coefficient-function differences are non-negligible. On the other hand, our 
results confirm that these quantities are very small, and thus relevant only 
when a high accuracy is required. Presently this condition is fulfilled 
only for the determination of the weak mixing angle $\theta_W$ from neutrino 
DIS to which we therefore turn now.

For this purpose one considers the so-called Paschos-Wolfenstein relation 
defined in terms of a ratio of neutral-current and charged-current cross 
sections for neutrino-nucleon DIS~\cite{Paschos:1973kj},
\beq
\label{eq:rminus}
R^{-} \; = \:\:  
\frac{\sigma(\nu_{\mu\,}N\rightarrow\nu_{\mu\,}X) \: - \: 
  \sigma(\bar \nu_{\mu\,}N\rightarrow\bar \nu_{\mu\,}X)}
{\sigma(\nu_{\mu\,}N\rightarrow\mu^-X) \: - \: 
  \sigma(\bar \nu_{\mu\,}N\rightarrow\mu^+X)} 
\:\: .
\eeq
$R^{-}$ directly measures $\,\sin^2 \theta_W$ if the up and down valence quarks 
in the target carry equal momenta, and if the strange and heavy-quark sea
distributions are charge symmetric. At the lowest order of perturbative QCD 
one generally finds
\beq
\label{eq:rminusLO}
R^{-}_{\:\!\rm LO} \:\: = \:\: {1 \over 2} \: - \: \sin^2 \theta_W 
\:\: .
\eeq
The quantity (\ref{eq:rminus}) has attracted considerable attention in recent 
years due to a determination of $\sin^2 \theta_W$ by the NuTeV collaboration
\cite{Zeller:2001hh}: within the Standard Model their result is at variance 
with other measurements of this quantity~\cite{Yao:2006px}, see also Refs.~\cite
{Davidson:2001ji,McFarland:2003jw,Dobrescu:2003ta} for detailed discussions.
   
Beyond the leading order Eq.~(\ref{eq:rminusLO}) receives perturbative QCD 
corrections which involve the second moments of coefficient functions for 
the $\nu N - \bar\nu N$ neutral- and charged-current structure functions.%
\footnote{$\,$Specifically the ratio $R^{-}$ includes, besides all $\,\nu N 
- \bar\nu N$ CC coefficient functions, the neutral-current quantity 
$C_3^{\:\rm NC}$ which is equal to its charged-current counterpart 
$C_3^{\,\nu N - \bar\nu N}$ at the perturbative orders considered here.}
Armed with the results of Sections \ref{sec:2-loop} and \ref{sec:3-loop} we are
now able to finalize the corresponding $\as^{\,2}$ contribution for massless 
quarks \cite{McFarland:2003jw} and to present an accurate numerical result at 
order $\as^{\,3}$.
We denote by $q^- \equiv q-\bar{q}\,$ the second Mellin moments of the 
valence distributions of the flavours $q = u,\;d,\;s,\;\ldots\,$, 
\beq
  \label{eq:pdfmom}
  q^- \; = \; \int_0^1\! dx\; x \left( q(x) - \bar{q}(x) \right)\:\: .
\eeq
The QCD corrections to $R^{-}$ can be expanded in inverse powers of the 
dominant isoscalar combination $\,u^- + d^-$ of the parton distributions~~--~~%
recall that the measurements of this ratio are performed for (almost) isoscalar
targets. After inserting the expansion of the \MSb\ coefficient functions in 
powers of $\as\,$, the Paschos--Wolfenstein ratio Eq.~(\ref{eq:rminus}) can be 
written as
\bea
\label{eq:rminusNNNLO}
R^{-} &\! =& 
g_L^{\:\!2} - g_R^{\:\!2} 
  \; + \; \frac{u^- - d^- + c^- - s^-}{u^- + d^-} \: \Biggl( 
  3(g_{Lu}^{\:\!2} - g_{Ru}^{\:\!2}) + (g_{Ld}^{\:\!2} - g_{Rd}^{\:\!2}) 
\nn \\ & & \mbox{}
+ (g_L^{\:\!2} - g_R^{\:\!2}) \: \Biggl\{ 
  \,\frac{8}{9} \frac{\as}{\pi} 
  \; +\; \frac{\alpha_s^2}{\pi^2} \: \Biggl[ 
      \frac{15127}{1944} 
    - \frac{89}{81}\, \z2 
    + \frac{61}{27}\, \z3
    - \frac{32}{45}\, \zs2 
    - \frac{83}{162}\, \nf 
    \Biggr] 
\nn\\ & & \mbox{}
     +\; \frac{\as^3}{\pi^3} \Biggl[ 
       \frac{5175965}{52488}
     - \frac{356}{729}\, \z2
     - \frac{586}{27}\, \z3
     - \frac{128}{405}\, \zs2
     + \frac{190}{81}\, \z5
     - \frac{9062}{729}\, \nf
     + \frac{2}{3}\, \nf \z3
\nn\\ & & \mbox{}
     + \frac{226}{729}\, \nfs
     - \frac{1}{32}\, \delta c_{2,2}^{(3)}
     + \frac{1}{128}\, \delta c_{L,2}^{(3)}
    \Biggr] \Biggr\} \Biggr)
  \; + \; {\cal{O}} \left( (u^- + d^-)^{-2\,} \right)
  \; + \; {\cal{O}}\bigl(\alpha_s^4\bigr) \:\: .
\eea
Here the left- and right-handed weak couplings $\,g_{Lu}$, $g_{Ld}$, $g_{Ru\,}$ 
and $g_{Rd\,}$ are related to the weak mixing angle $\sin^2 \theta_W$ by
\beq
\label{eq:gLgRdef}
 g_L^{\:\!2} \;\equiv\; g_{Lu}^{\:\!2} + g_{Ld}^{\:\!2} 
 \;=\; \frac{1}{2}-\sin^2\theta_W+\frac{5}{9}\sin^4\theta_W
\:\: ,\qquad
 g_R^{\:\!2} \;\equiv\; g_{Ru}^{\:\!2} + g_{Rd}^{\:\!2} 
 \;=\;  \frac{5}{9}\sin^4\theta_W 
\:\: .
\eeq
Beyond the tree level, of course, these relations receive electroweak radiative
corrections, see, e.g., Ref.~\cite{Diener:2005me}. Eq.~(\ref{eq:rminusNNNLO}) 
shows the well-known fact that the relation (\ref{eq:rminusLO}) receives 
corrections if the parton content of the target includes an isotriplet 
component, $u^-\not=d^-$, or a quark sea with a $C$-odd component, 
$s^-\not=0\,$ or $\,c^-\not=0$. 
Notice also that perturbative QCD only affects these corrections. 

The exact second-order contribution in Eq.~(\ref{eq:rminusNNNLO}) differs 
from the result in Ref.~\cite{McFarland:2003jw} where the function 
$\,\delta\:\!  c_{L}^{(2)}(x)$ of Eq.~(\ref{eq:dcLqq2}) was not included.
The third-order corrections can now be completed in a numerical form, using
our approximations (\ref{eq:dc3mom2}) for the second moments of
$\,\delta\:\! c_{2,L}^{(3)}(x)$. For $\nf = 4\,$ flavours (and disregarding
electroweak corrections) we obtain   
\bea
\label{eq:rminus-numbers}
R^{-} &\! = &
 \frac{1}{2} - \sin^2\theta_W 
  \:\: + \:\: \frac{u^- - d^- + c^- - s^-}{u^- + d^-} \: \Bigg\{
  1 - \frac{7}{3}\:\sin^2\theta_W  
  \; + \; \left( \frac{1}{2} - \sin^2\theta_W \right) \cdot
\nn \\ & & \mbox{} 
  \frac{8}{9} \frac{\alpha_s}{\pi} \left[ \,
    1
  + 1.689\,\as
  + (3.661 \pm 0.002)\,\as^2 \,
  \right]
  \Biggr\}
  \; + \; {\cal{O}} \left( (u^- + d^-)^{-2\,} \right)
  \; + \; {\cal{O}}\bigl(\alpha_s^4\bigr) 
\:\: . \quad
\eea
The perturbation series in the square brackets appears reasonably well 
convergent for relevant values of the strong coupling constant, with the known
terms reading, e.g., 1 + 0.42 + 0.23 for $\as = 0.25$. Thus the $\as^2$ and 
$\as^3$ contributions correct the NLO estimate by 65\% in this case. On the 
other hand, due to the small prefactor of this expansion, the new third-order 
term increases the complete curved bracket in Eq.~(\ref{eq:gLgRdef}) by only 
about 1\%, which can therefore by considered as the new uncertainty of this 
quantity due to the truncation of the perturbative expansion. Consequently
previous NLO estimates of the effect of, for instance, the (presumably mainly 
non-perturbative, see Refs.~\cite{Catani:2004nc,Lai:2007dq,Thorne:2007bt}) 
charge asymmetry of the strange sea remain practically unaffected by 
higher-order corrections to the coefficient functions. 
%
%
\setcounter{equation}{0}
\section{Summary}
\label{sec:summary}
%
%
In this article we have presented new results for the coefficient functions of
inclusive charged-current DIS in the framework of massless perturbative QCD.
We have filled a gap in the two-loop literature by writing down the 
corresponding difference $\,\delta\:\! c^{(2)}_L(x)$ of the $\,\nu p + 
\bar\nu p\,$ and \mbox{$\,\nu p - \bar\nu p\,$} structure functions $\,F_{L\,}$.
Our main results are the lowest five (even- or odd-integer) Mellin moments of 
the third-order corrections $\,\delta\:\! c^{(3)}_a(x)$ for all three structure
functions $F_{a\: =\: 2,\:3,\:L}$ and approximations in Bjorken-$x$ space based 
on these moments  which are applicable down to at least $x \lsim 10^{-2}$.
As a byproduct we have calculated the related third-order coefficient-function 
correction to the Gottfried sum rule in photon-exchange DIS.

All our third-order results are proportional to the `non-planar' colour factor
$C_A-2\,C_F$, thus confirming a conjecture by Broadhurst, Kataev and Maxwell on 
the $1/N_c^{\,2}$ suppression of these coefficient-function differences in the
limit of a large number of colours $N_c$. Numerically our $\as^3$ corrections 
prove relevant in particular for $F_{\:\!2}$ and $F_L$ wherever the differences 
of the $\,\nu p + \bar\nu p\,$ and $\,\nu p - \bar\nu p\,$ coefficient functions
are not negligible. We have employed the above results to derive the second- and 
third-order QCD corrections to the Paschos-Wolfenstein ratio $R^-$ used to 
determine the weak mixing angle from neutrino deep-inelastic scattering. The
uncertainty due to uncalculated higher-order coefficient functions has been 
reduced to a level amply sufficient for the foreseeable future, i.e., 1\% for the
coefficient-function factor multiplying the quark-distribution asymmetries.

{\sc Form} files and {\sc Fortran} subroutines with our results can be obtained 
from the preprint server {\tt http://arXiv.org} by downloading the source of 
this article.  Furthermore they are available from the authors upon request.
%
%
\subsection*{Note added}
While this article was finalized, the 11-th moments of the functions 
$\,\delta\:\! c^{(3)}_2(x)$ and $\,\delta\:\! c^{(3)}_L(x)$ have been 
computed~\cite{Rogalprep}. Both results fall into the bands generated by the 
respective $x$-space approximations in Section~\ref{sec:3-loop}, thus confirming
the reliability of these uncertainty estimates.
%
%
\subsection*{Acknowledgements}
We would like to thank S.~Alekhin, D.~Broadhurst and A.~Kataev for stimulating 
discussions.
The work of S.M. and M.R. has been supported by the Helmholtz Gemeinschaft 
under contract VH-NG-105 and in part by the Deutsche Forschungsgemeinschaft in 
Sonderforschungs\-be\-reich/Transregio~9. During the final stage of this 
research A.V. enjoyed the hospitality of the Instituut-Lorentz of Leiden
University.
%
%
\section*{Appendix}
\renewcommand{\theequation}{A.\arabic{equation}}
\setcounter{equation}{0}
%
%
Here we present the analytic expressions for the Mellin-space 
coefficient-function differences $\delta c_{a,N}^{(3)}$ which were given 
numerically in Eqs.~(\ref{eq:dc2ns}) -- (\ref{eq:dc3ns}). We use the 
notations and conventions as specified at the beginning of 
Section~\ref{sec:2-loop} and above Eq.~(\ref{eq:dc2qq2} ).

The first moment of $\delta c_{2}^{(3)}(x)$ has been written down in Eq.~(\ref
{eq:dc2q1}) above. The remaining known moments of this quantity are given by
\bea
\label{eq:dc2q3}
  \delta c_{2,3}^{(3)} &\! =\! &
        \cf \* \cfas  \*  \Biggl(
           {1805677051 \over 466560}
          - {2648 \over 9} \* \z5
          + {10093427 \over 810} \* \z3
          - {1472 \over 3} \* \zs3
          - {7787113 \over 1944} \* \z2
\nonumber\\
& &\mbox{\hspn}\qquad\qquad
          + {55336 \over 9} \* \z2 \* \z3
          - {378838 \over 45} \* \zs2
          - {8992 \over 63} \* \zt2
          \Biggr)
\nonumber\\
& &\mbox{\hspn}
       + \cfs \* \cfa  \*  \Biggl(
          - {5165481803 \over 1399680}
          + {40648 \over 9} \* \z5
          - {9321697 \over 810} \* \z3
          + {1456 \over 3} \* \zs3
          + {8046059 \over 1944} \* \z2
 \nonumber\\
& &\mbox{\hspn}\qquad\qquad
          - 4984 \* \z2 \* \z3
          + {798328 \over 135} \* \zs2
          - {56432 \over 315} \* \zt2
          \Biggr)
\\
& &\mbox{\hspn}
       + \nf \* \cf \* \cfa \*  \Biggl(
           {20396669 \over 116640}
          - {1792 \over 9} \* \z5
          + {405586 \over 405} \* \z3
          - {139573 \over 486} \* \z2
          + {1408 \over 9} \* \z2 \* \z3
          - {50392 \over 135} \* \zs2
          \Biggr)
\nonumber\; ,
\quad
\\[2mm]
\label{eq:dc2q5}
  \delta c_{2,5}^{(3)} & = &
        \cf \* \cfas  \*  \Biggl(
           {18473631996593 \over 3827250000}
          - {17584 \over 45} \* \z5
          + {149815672 \over 7875} \* \z3
          - {1472 \over 3} \* \zs3
 \nonumber\\
& &\mbox{\hspn}\qquad\qquad
          - {291199027 \over 50625} \* \z2
          + {330416 \over 45} \* \z2 \* \z3
          - {2577928 \over 225} \* \zs2
          - {8992 \over 63} \* \zt2
          \Biggr)
\nonumber\\
& &\mbox{\hspn}
       + \cfs \* \cfa  \*  \Biggl(
          - {16016244428419 \over 3827250000}
          + {47560 \over 9} \* \z5
          - {1270840912 \over 70875} \* \z3
          + {1456 \over 3} \* \zs3
 \nonumber\\
& &\mbox{\hspn}\qquad\qquad
          + {1321405949 \over 202500} \* \z2
          - {89128 \over 15} \* \z2 \* \z3
          +{26658224 \over 3375} \* \zs2
          - {56432 \over 315} \* \zt2
          \Biggr)
\nonumber\\
& &\mbox{\hspn}
       + \nf \* \cf \* \cfa  \*  \Biggl(
           {181199822513 \over 765450000}
          - {1792 \over 9} \* \z5
          + {6514448 \over 4725} \* \z3
          - {1652773 \over 3375} \* \z2
 \nonumber\\
& &\mbox{\hspn}\qquad\qquad
          + {1408 \over 9} \* \z2 \* \z3
          - {11888 \over 27} \* \zs2
          \Biggr)
\; ,
\quad
\\[2mm]
\label{eq:dc2q7}
  \delta c_{2,7}^{(3)} & = &
       \cf \* \cfas  \*  \Biggl(
           {177036089007294328733 \over 32934190464000000}
          - {27248 \over 63} \* \z5
          + {65397081433 \over 2646000} \* \z3
 \nonumber\\
& &\mbox{\hspn}\qquad\qquad
          - {1472 \over 3} \* \zs3
          - {340303364748629 \over 46675440000} \* \z2
          + {2563996 \over 315} \* \z2 \* \z3
 \nonumber\\
& &\mbox{\hspn}\qquad\qquad
          - {4570738447 \over 330750} \* \zs2
          - {8992 \over 63} \* \zt2
          \Biggr)
\nonumber\\
& &\mbox{\hspn}
       + \cfs \* \cfa  \*  \Biggl(
          - {213694072871074531 \over 45177216000000}
          + {1821772 \over 315} \* \z5
          - {438487320707 \over 18522000} \* \z3
 \nonumber\\
& &\mbox{\hspn}\qquad\qquad
          + {1456 \over 3} \* \zs3
          + {418808510000479 \over 46675440000} \* \z2
          - {2071492 \over 315} \* \z2 \* \z3
 \nonumber\\
& &\mbox{\hspn}\qquad\qquad
          + {6241478743 \over 661500} \* \zs2
          - {56432 \over 315 } \* \zt2
          \Biggr)
\nonumber\\
& &\mbox{\hspn}
       + \nf \* \cf \* \cfa  \*  \Biggl(
           {38079608000704561 \over 117622108800000}
          - {1792 \over 9} \* \z5
          + {22115039 \over 13230} \* \z3
 \nonumber\\
& &\mbox{\hspn}\qquad\qquad
          - {113587875043 \over 166698000} \* \z2
          + {1408 \over 9} \* \z2 \* \z3
          - {2296328 \over 4725} \* \zs2
          \Biggr)
\; ,
\quad
\\[2mm]
\label{eq:dc2q9}
  \delta c_{2,9}^{(3)} & = &
       \cf \* \cfas  \*  \Biggl(
           {5676515460744370321603 \over 1000376035344000000}
          - {25664 \over 63} \* \z5
          + {11165079556403 \over 375070500} \* \z3
 \nonumber\\
& &\mbox{\hspn}\qquad\qquad
          - {1472 \over 3} \* \zs3
          - {8178803099431493 \over 945177660000} \* \z2
          + {1648352 \over 189} \* \z2 \* \z3
 \nonumber\\
& &\mbox{\hspn}\qquad\qquad
          - {23488033336 \over 1488375} \* \zs2
          - {8992 \over 63} \* \zt2
          \Biggr)
\nonumber\\
& &\mbox{\hspn}
       + \cfs \* \cfa  \*  \Biggl(
          - {32102287673972370020989 \over 6002256212064000000}
          + {1162796 \over 189} \* \z5
          - {89153747611 \over 3087000} \* \z3
 \nonumber\\
& &\mbox{\hspn}\qquad\qquad
          + {1456 \over 3} \* \zs3
          + {342078312478997 \over 30005640000} \* \z2
          - {1332820 \over 189} \* \z2 \* \z3
 \nonumber\\
& &\mbox{\hspn}\qquad\qquad
          + {3187232017 \over 297675} \* \zs2
          - {56432 \over 315} \* \zt2
          \Biggr)
\nonumber\\
& &\mbox{\hspn}
       + \nf \* \cf \* \cfa  \*  \Biggl(
           {21832132134852204299 \over 52400649470400000}
          - {1792 \over 9} \* \z5
          + {6271692134 \over 3274425} \* \z3
 \nonumber\\
& &\mbox{\hspn}\qquad\qquad
          - {1931824297943 \over 2250423000} \* \z2
          + {1408 \over 9} \* \z2 \* \z3
          - {164116 \over 315} \* \zs2
          \Biggr)
\; .
\end{eqnarray}
The corresponding lowest five odd-integer moments for the longitudinal
structure function read 
\begin{eqnarray}
\label{eq:dcLq1}
  \delta c_{L,1}^{(3)} & = &
        \cf \*\cfas \* \Biggl(
         { 21977 \over 9}
          -{ 608 \over3}\*\z5
          -{ 2648 \over9}\*\z3
          -{ 3068 \over 9}\*\z2
          - 448\*\z2\*\z3
          - 336\*\zs2
          \Biggr)
\nonumber\\
& &\mbox{\hspn}
       +  \cfs \* \cfa \* \Biggl(
          -{ 17819 \over 9}
          -{ 1568 \over 3}\*\z5
          + {5648 \over 9}\*\z3
          + {1376 \over 9}\*\z2
          + 288\*\z2\*\z3
          + {2304 \over 5}\*\zs2
          \Biggr)
\nonumber\\
& &\mbox{\hspn}
       +  \nf \* \cf \* \cfa \* \Biggl(
          { 1366 \over 9}
          -{ 496 \over 9}\*\z3
          -{ 328 \over 9}\*\z2
          -{ 224 \over 15}\*\zs2
          \Biggr)
\; ,
\quad
\\[2mm]
\label{eq:dcLq3}
  \delta c_{L,3}^{(3)} & = &
      \cf \* \cfas  \*  \Biggl(
          - {12350749 \over 19440}
          + 352 \* \z5
          + {52516 \over 45} \* \z3
          +{ 47 \over 27} \* \z2
          + 96 \* \z2 \* \z3
          - {7544 \over 15} \* \zs2
          \Biggr)
\nonumber\\
& &\mbox{\hspn}
       + \cfs \* \cfa  \*  \Biggl(
           {10152961 \over 12960}
          - 368 \* \z5
          - {16412 \over 15} \* \z3
          - 242 \* \z2
          + 144 \* \z2 \* \z3
          +{ 1168 \over 3} \* \zs2
          \Biggr)
\nonumber\\
& &\mbox{\hspn}
       + \nf \* \cf \* \cfa  \*  \Biggl(
          - {16757 \over 1620}
          +{ 2936 \over 45} \* \z3
          - {16 \over 9} \* \z2
          - {368 \over 15} \* \zs2
          \Biggr)
\; ,
\quad
\\[2mm]
\label{eq:dcLq5}
  \delta c_{L,5}^{(3)} & = &
       \cf \* \cfas  \*  \Biggl(
          - {735306721 \over 17010000}
          - {1888 \over 3} \* \z5
          + {558244 \over 315} \* \z3
          - {442783 \over 675} \* \z2
 \nonumber\\
& &\mbox{\hspn}\qquad\qquad
          + 448 \* \z2 \* \z3
          - {4160 \over 9} \* \zs2
          \Biggr)
\nonumber\\
& &\mbox{\hspn}
       + \cfs \* \cfa  \*  \Biggl(
           {6741265367 \over 10206000}
          - {736 \over 3} \* \z5
          - {1285168 \over 945} \* \z3
          + {51493 \over 405} \* \z2
 \nonumber\\
& &\mbox{\hspn}\qquad\qquad
          + 96 \* \z2 \* \z3
          + {69608 \over 225} \* \zs2
          \Biggr)
\nonumber\\
& &\mbox{\hspn}
       + \nf \* \cf \* \cfa  \*  \Biggl(
          - {2107157 \over 255150}
          + {8816 \over 105} \* \z3
          - {11992 \over 405} \* \z2
          - {736 \over 45} \* \zs2
          \Biggr)
\; ,
\quad
\\[2mm]
\label{eq:dcLq7}
  \delta c_{L,7}^{(3)} & = &
      \cf \* \cfas  \*  \Biggl(
           {354522585410107 \over 666792000000}
          - 1408 \* \z5
          + {47266403 \over 23625} \* \z3
          - {1095179473 \over 945000} \* \z2
 \nonumber\\
& &\mbox{\hspn}\qquad\qquad
          + 752 \* \z2 \* \z3
          - {147056 \over 375} \* \zs2
          \Biggr)
\nonumber\\
& &\mbox{\hspn}
       + \cfs \* \cfa  \*  \Biggl(
           {11388456807174161 \over 28005264000000}
          - 184 \* \z5
          - {176925641 \over 132300} \* \z3
          + {4569363329 \over 13230000 }\* \z2
 \nonumber\\
& &\mbox{\hspn}\qquad\qquad
          + 72 \* \z2 \* \z3
          + {663878 \over 2625} \* \zs2
          \Biggr)
\nonumber\\
& &\mbox{\hspn}
       + \nf \* \cf \* \cfa  \*  \Biggl(
           {369546282989 \over 50009400000}
          + {124282 \over 1575} \* \z3
          - {220747 \over 5250} \* \z2
          - {184 \over 15} \* \zs2
          \Biggr)
\; ,
\quad
\\[2mm]
\label{eq:dcLq9}
  \delta c_{L,9}^{(3)} & = &
      \cf \* \cfas  \*  \Biggl(
           {1346454911003496947 \over 1323248724000000}
          - {10528 \over 5 }\* \z5
          + {13247918 \over 6125} \* \z3
 \nonumber\\
& &\mbox{\hspn}\qquad\qquad
          - {325373958827 \over 208372500} \* \z2
          + {5184 \over 5} \* \z2 \* \z3
          - {296736 \over 875} \* \zs2
          \Biggr)
\nonumber\\
& &\mbox{\hspn}
       + \cfs \* \cfa  \*  \Biggl(
           {17693872049573089 \over 73513818000000}
          - {736 \over 5} \* \z5
          - {125991917 \over 99225} \* \z3
 \nonumber\\
& &\mbox{\hspn}\qquad\qquad
          + {10496201057 \over 23152500} \* \z2
          + {288 \over 5} \* \z2 \* \z3
          + {1688888 \over 7875} \* \zs2
          \Biggr)
\nonumber\\
& &\mbox{\hspn}
       + \nf \* \cf \* \cfa  \*  \Biggl(
           {23852323249607 \over 1444021425000}
          + {1249264 \over 17325 }\* \z3
          - {1542176 \over 33075} \* \z2
          - {736 \over 75} \* \zs2
          \Biggr)
\; .
\end{eqnarray}
Finally the analytic expressions for the moments $\delta c_{2,N}^{(3)}$ in
Eq.~(\ref{eq:dc3ns}) are
\begin{eqnarray}
\label{eq:dc3q2}
  \delta c_{3,2}^{(3)} & = &
       \cf \* \cfas  \*  \Biggl(
           -{840949 \over 243}
          + {9344 \over 9} \* \z5
          - {650360 \over 81 }\* \z3
          + {1472 \over 3} \* \zs3
          + {712328 \over 243} \* \z2
 \nonumber\\
& &\mbox{\hspn}\qquad\qquad
          - {47920 \over 9} \* \z2 \* \z3
          + {30416 \over 5} \* \zs2
          + {8992 \over 63} \* \zt2
          \Biggr)
\nonumber\\
& &\mbox{\hspn}
       + \cfs \* \cfa  \*  \Biggl(
           {15979879 \over 4374}
          - {38416 \over 9 }\* \z5
          + {580504 \over 81} \* \z3
          - {1456 \over 3} \* \zs3
          - {742390 \over 243} \* \z2
 \nonumber\\
& &\mbox{\hspn}\qquad\qquad
          + 4368 \* \z2 \* \z3
          - {577264 \over 135} \* \zs2
          + {56432 \over 315} \* \zt2
          \Biggr)
 \\
& &\mbox{\hspn}
       + \nf \* \cf \* \cfa  \*  \Biggl(
          -{119522 \over 729}
          + {1792 \over 9} \* \z5
          - {57128 \over 81} \* \z3
          + {46112 \over 243 }\* \z2
          - {1408 \over 9 }\* \z2 \* \z3
          + {40024 \over 135} \* \zs2
          \Biggr)
\nonumber\; ,
\quad
\\[2mm]
\label{eq:dc3q4}
  \delta c_{3,4}^{(3)} & = &
       \cf \* \cfas  \*  \Biggl(
          -{21230721185377 \over 4374000000}
          + {23704 \over 45} \* \z5
          - {292322783 \over 20250} \* \z3
          + {1472 \over 3} \* \zs3
 \nonumber\\
& &\mbox{\hspn}\qquad\qquad
          + {5644168873 \over 1215000} \* \z2
          - {100792 \over 15} \* \z2 \* \z3
          + {6477802 \over 675} \* \zs2
          + {8992 \over 63} \* \zt2
          \Biggr)
\nonumber\\
& &\mbox{\hspn}
       + \cfs \* \cfa  \*  \Biggl(
           {19991706724601 \over 4374000000}
          - 5208 \* \z5
          + {272933467 \over 20250} \* \z3
          - {1456 \over 3} \* \zs3
 \nonumber\\
& &\mbox{\hspn}\qquad\qquad
          - {6307524619 \over 1215000} \* \z2
          + {253064 \over 45} \* \z2 \* \z3
          - {2500616 \over 375} \* \zs2
          + {56432 \over 315} \* \zt2
          \Biggr)
\nonumber\\
& &\mbox{\hspn}
       + \nf \* \cf \* \cfa  \*  \Biggl(
           -{15339664501 \over 72900000}
          + {1792 \over 9} \* \z5
          - {755894 \over 675} \* \z3
          + {21942049 \over 60750} \* \z2
 \nonumber\\
& &\mbox{\hspn}\qquad\qquad
          - {1408 \over 9} \* \z2 \* \z3
          + {53128 \over 135} \* \zs2
          \Biggr)
\; ,
\quad
\\[2mm]
\label{eq:dc3q6}
  \delta c_{3,6}^{(3)} & = &
       \cf \* \cfas  \*  \Biggl(
          -{172761364527374293 \over 32162295375000}
          + {21200 \over 63} \* \z5
          - {3380925064 \over 165375} \* \z3
          + {1472 \over 3} \* \zs3
 \nonumber\\
& &\mbox{\hspn}\qquad\qquad
          + {147865501939 \over 24310125} \* \z2
          - {2395856 \over 315} \* \z2 \* \z3
          + {75351016 \over 6125} \* \zs2
          + {8992 \over 63} \* \zt2
          \Biggr)
\nonumber\\
& &\mbox{\hspn}
       + \cfs \* \cfa  \*  \Biggl(
           {313157547783370669 \over 64324590750000}
          - {1810712 \over 315 }\* \z5
          + {67828543996 \over 3472875} \* \z3
          - {1456 \over 3} \* \zs3
 \nonumber\\
& &\mbox{\hspn}\qquad\qquad
          - {3604225183081 \over 486202500} \* \z2
          + {667864 \over 105} \* \z2 \* \z3
          - {1397140016 \over 165375} \* \zs2
          + {56432 \over 315} \* \zt2
          \Biggr)
\nonumber\\
& &\mbox{\hspn}
       + \nf \* \cf \* \cfa  \*  \Biggl(
           -{503591542653161 \over 1837845450000}
          + {1792 \over 9} \* \z5
          - {16004944 \over 11025} \* \z3
          + {23420609 \over 42875} \* \z2
 \nonumber\\
& &\mbox{\hspn}\qquad\qquad
          - {1408 \over 9} \* \z2 \* \z3
          + {2135888 \over 4725} \* \zs2
          \Biggr)
\; ,
\quad
\\[2mm]
\label{eq:dc3q8}
  \delta c_{3,8}^{(3)} & = &
       \cf \* \cfas  \*  \Biggl(
           -{45882775286477927067311 \over 8003008282752000000}
          + {22640 \over 63} \* \z5
          - {38829577931303 \over 1500282000} \* \z3
 \nonumber\\
& &\mbox{\hspn}\qquad\qquad
           +{1472 \over 3} \* \zs3
          + {5677110453154657 \over 756142128000} \* \z2
          - {7858468 \over 945} \* \z2 \* \z3
 \nonumber\\
& &\mbox{\hspn}\qquad\qquad
          + {43146817871 \over 2976750} \* \zs2
          + {8992 \over 63} \* \zt2
          \Biggr)
\nonumber\\
& &\mbox{\hspn}
       + \cfs \* \cfa  \*  \Biggl(
           {126830527574348410837327 \over 24009024848256000000}
          - {5795836 \over 945} \* \z5
          + {4175977929883 \over 166698000} \* \z3
 \nonumber\\
& &\mbox{\hspn}\qquad\qquad
          - {1456 \over 3} \* \zs3
          - {194854342276579 \over 20003760000} \* \z2
          + {6509876 \over 945} \* \z2 \* \z3
 \nonumber\\
& &\mbox{\hspn}\qquad\qquad
          - {58805551031 \over 5953500} \* \zs2
          + {56432 \over 315} \* \zt2
          \Biggr)
\nonumber\\
& &\mbox{\hspn}
       + \nf \* \cf \* \cfa  \*  \Biggl(
           -{3380190329263337489 \over 9527390812800000}
          + {1792 \over 9} \* \z5
          - {1026540911 \over 595350} \* \z3
 \nonumber\\
& &\mbox{\hspn}\qquad\qquad
          + {3263620615369 \over 4500846000} \* \z2
          - {1408 \over 9} \* \z2 \* \z3
          + {778496 \over 1575} \* \zs2
          \Biggr)
\; ,
\quad
\\[2mm]
\label{eq:dc3q10}
  \delta c_{3,10}^{(3)} & = &
       \cf \* \cfas  \*  \Biggl(
           -{2924815993615556996346598663 \over 483334682604559632000000}
          + {210944 \over 385} \* \z5
 \nonumber\\
& &\mbox{\hspn}\qquad\qquad
          - {3080312718428437 \over 99843767100} \* \z3
          + {1472 \over 3} \* \zs3
          + {13720175530646448109 \over 1537594013340000} \* \z2
 \nonumber\\
& &\mbox{\hspn}\qquad\qquad
          - {8455904 \over 945} \* \z2 \* \z3
          + {3568998808 \over 218295} \* \zs2
          + {8992 \over 63} \* \zt2
          \Biggr)
\nonumber\\
& &\mbox{\hspn}
       + \cfs \* \cfa  \*  \Biggl(
           {61916581373996975119251441821 \over 10633363017300311904000000} 
          - {66873844 \over 10395} \* \z5
 \nonumber\\
& &\mbox{\hspn}\qquad\qquad
           + {30055925797598243 \over 998437671000} \* \z3
          - {1456 \over 3} \* \zs3
          - {3186598606475201011 \over 263587545144000} \* \z2
 \nonumber\\
& &\mbox{\hspn}\qquad\qquad
          + {75900812 \over 10395} \* \z2 \* \z3
          - {1995571648453 \over 180093375} \* \zs2
          + {56432 \over 315} \* \zt2
          \Biggr)
\nonumber\\
& &\mbox{\hspn}
       + \nf \* \cf \* \cfa  \*  \Biggl(
           -{339629926756418877268603 \over 767197908896126400000}
          +{1792 \over 9} \* \z5
          - {70469642338 \over 36018675} \* \z3
 \nonumber\\
& &\mbox{\hspn}\qquad\qquad
          + {2677118231310293 \over 2995313013000} \* \z2
          - {1408 \over 9} \* \z2 \* \z3
          + {1015276 \over 1925} \* \zs2
          \Biggr)
\; .
\end{eqnarray}
%
\end{document}